\documentclass[11pt]{article}
\usepackage{amsmath,amssymb,color,graphics,epsfig,cite}

\usepackage[unicode=false,
bookmarks=true,bookmarksnumbered=false,bookmarksopen=false,
breaklinks=false,pdfborder={0 0 1},backref=false,colorlinks=false]
{hyperref}
\usepackage{amsfonts}

\newcommand{\hoch}[1]{$\, ^{#1}$}

\newcommand{\be}{\begin{equation}}
\newcommand{\ee}{\end{equation}}
\newcommand{\bea}{\setlength\arraycolsep{2pt} \begin{eqnarray}}
\newcommand{\eea}{\end{eqnarray}}
\newcommand{\nn}{\nonumber}

\def\crampest{\medmuskip = 1mu plus 1mu minus 1mu}

\def\ft#1#2{{\textstyle{\frac{\scriptstyle #1}{\scriptstyle #2} } }}
\def\fft#1#2{{\frac{#1}{#2}}}

\def\0{{\sst{(0)}}}
\def\1{{\sst{(1)}}}
\def\2{{\sst{(2)}}}
\def\3{{\sst{(3)}}}
\def\4{{\sst{(4)}}}
\def\5{{\sst{(5)}}}
\def\6{{\sst{(6)}}}
\def\7{{\sst{(7)}}}
\def\8{{\sst{(8)}}}
\def\sst#1{{\scriptscriptstyle #1}}
\def\oneone{\rlap 1\mkern4mu{\rm l}}

\begin{document}

\begin{center}
{\Large {\bf Instanton Moduli, Topology\\ and the Bosonic/Heterotic String Origins }}

\vspace{20pt}

Liang Ma\hoch{1} and H. L\"{u}\hoch{1,2}

\vspace{10pt}

{\it \hoch{1}Center for Joint Quantum Studies, Department of Physics,\\
School of Science, Tianjin University, Tianjin 300350, China }

\medskip
		
{\it \hoch{2}The International Joint Institute of Tianjin University, Fuzhou,\\ Tianjin University, Tianjin 300350, China}

\vspace{40pt}

\underline{ABSTRACT}
\end{center}

We construct new supersymmetric NS(-1)-branes supported by the $SU(2)\times SU(2)$ 't Hooft instantons in ${\cal N}=1$, $D=4$ supergravity. We show that although the magnetic 3-form string flux is invariant under the variation of instanton moduli, the string-frame metric can change topology, giving rise to either $\mathbb R^4$ or the new $\mathbb R\times S^3$ bolt topologies. We uplift the solutions to both heterotic and bosonic strings, where the instanton moduli become parameters of pure geometry. We find that the Killing spinors exist in both theories. The transformation of the Killing spinors from one theory to the other is consistent with the recently proposed bosonic/heterotic string duality. We further construct new BPS solutions in the $D=6$ Salam-Sezgin model, employing the technique developed in the paper.

\vfill {\footnotesize maliang0@tju.edu.cn\ \ \ mrhonglu@gmail.com}


\thispagestyle{empty}
\pagebreak



\section{Introduction}

In string theory, the dynamics of the fundamental string are governed by the worldsheet action \cite{Goto:1971ce,Polyakov:1981rd}, while the low-energy theory is (super)gravity that includes the metric,  a dilaton and 2-form potential. In $D=10$ dimensions, the supergravity counterpart of the fundamental string is the string soliton solution \cite{Dabholkar:1990yf}. More precisely, the soliton is not an exact supergravity solution; it is singular and requires the string action as its source. This is in fact generally the case for supergravity $p$-branes, including the magnetic dual 5-brane of the fundamental string \cite{Duff:1994an}. There exist only a few counterexamples such as M5-brane \cite{Gueven:1992hh} and D3-brane \cite{Horowitz:1991cd,Duff:1991pea}, where the radial coordinate $r$ can extend to the negative region which is an identical copy of the exterior \cite{Gibbons:1994vm}, analogous to the wormhole geometry.

Smooth $p$-brane solutions in supergravity without external source, can also be constructed by utilizing the transgression terms in a relevant form field strength. This leads to modified Bianchi identity that can supply a smooth source for the $p$-brane charge. The best-known example is perhaps the heterotic 5-brane \cite{Strominger:1990et}, supported by the magnetic charge of 3-form $H_\3$, satisfying the Bianchi identity
\be
dH_\3= \ft14 F^I_\2\wedge F_\2^I\,,
\ee
where $F^I_\2$'s are certain Yang-Mills 2-form field strengths, associated with gauge group either $E_8\times E_8$ or $SO(32)$. In $D=10$, the transverse space of a 5-brane is Euclidean space $\mathbb R^4$, which can support $SU(2)$ 't Hooft instanton solutions. These gauge instantons provide a smooth source for the NS-NS 5-brane, rending the solution is completely regular and eliminating the need for external sources \cite{Strominger:1990et}. It turns out that the resolution of singularities in six-dimensional gauge dyonic string is closely related to the absence of the phase transition of heterotic string compactified on K3 \cite{Duff:1996cf}. Brane resolution via transgression \cite{Cvetic:2000mh} can also be achieved through abelian form fields, provided that the transverse space is a suitable smooth Ricci-flat space with reduced holonomy. The best-known such example is perhaps the Klebanov-Strassler resolved D3-brane \cite{Klebanov:2000hb}.

In four dimensions, the magnetic dual of the fundamental string solution \cite{Greene:1989ya} is the ``-1''-brane, for which the entire ``spacetime'' is transverse and hence Euclidean. Such a solution is also referred to as an instanton. A notable example is the D-instanton, or D(-1)-brane associated with the R-R axion of type IIB string \cite{Bergshoeff:1998ry}. In this paper, To avoid the confusion between this use of the term ``instanton'' and Yang-Mills instantons, we shall refer to the Euclidean solution carrying the NS-NS 3-form flux as the ``NS(-1)''-brane.  The purpose of this work is to study these NS(-1)-branes in $D=4$, ${\cal N}=1$ supergravity coupled to a tensor multiplet and $SU(2)\times SU(2)$ Yang-Mills multiplets. The tensor multiplet is Hodge dual to a complex scalar multiplet. We examine how the 't Hooft Yang-Mills instantons affect the geometry and topology of the NS(-1)-branes. 

${\cal N}=1$, $D=4$ supergravity can be obtained from the heterotic string compactified on a Calabi-Yau manifold, with the ten-dimensional Yang-Mills fields descending directly to four dimensions. In this paper, we consider a very different origin: a nontrivial reduction on warped internal $\mathbb R\times T^{1,1}$ space, in which the $SU(2)\times SU(2)$ Yang-Mills gauge fields emerge from the isometry group of the $T^{1,1}$ space \cite{Ma:2025mvo}. This reduction ansatz arises from the recently proposed bosonic/heterotic duality, which relates the heterotic string theory with a linear dilaton along $\mathbb R$ to a noncritical bosonic string with conformal anomaly term at the string T-dualty fix point \cite{Ma:2025mvo,Lu:2025nyo}. The bosonic sector of the $D=4$ theory can be obtained from the simpler $S^3\times S^3$ reduction of the bosonic string theory. Since the bosonic string theory admits Killing spinor equations even though it is not supersymmetric \cite{Lu:2011zx,Liu:2012jra,Lu:2011ku}, this duality provides a mechanism for constructing new BPS solutions via the simpler bosonic string theory \cite{Lu:2025nyo}. In this paper, we construct NS(-1)-branes supported by the $SU(2)\times SU(2)$ 't Hooft instantons and then identity their origins in both heterotic and bosonic strings, which exhibit rich geometric and topological structures. We shall construct their Killing spinors in all respective theories.

The paper is organized as follows. In section 2, we introduce the ${\cal N}=1$, $D=4$ supergravity coupled to $SU(2)\times SU(2)$ Yang-Mills fields. We construct the NS(-1)-brane solution and study its geometry in the string frame. In particular, we examine how the Yang-Mills instanton affects the topology of the NS(-1)-brane. In section 3, we uplift the solutions to the bosonic and heterotic strings in $D=10$. We show that Killing spinors exist in all cases, providing support for the proposed bosonic/heterotic duality. In section 4, we observe that ${\cal N}=1$, $D=4$ supergravity with a single copy of $SU(2)$ Yang-Mills can be obtained via $S^2$ reduction \cite{Gibbons:2003gp} of the ${\cal N}=(1,0)$, $D=6$ Salam-Sezgin (SS) model \cite{Salam:1984cj}. This allows us to construct new BPS solutions in the SS model \cite{Salam:1984cj}. We conclude the paper in section 5. In appendix A, we provide some useful formulae related to $S^3$ and its embedding of $SU(2)$ Yang-Mills.

\section{Yang-Mills instanton moduli and NS(-1)-brane creation}

We begin with ${\cal N}=1$, $D=4$ supergravity coupled to one tensor multiplet and an $SU(2)\times SU(2)$ Yang-Mills multiplet. The bosonic sector consists of the metric, the tensor multiplet  $(\phi, B_\2)$ and two copies of $SU(2)$ Yang-Mills fields $(A^I, \widetilde A^{\tilde I})$. The Lagrangian 4-form is
\bea
\mathcal{L}_{4}=R\, {*\oneone}-\ft{1}{2} {* d\phi}\wedge d\phi-\ft{1}{2}e^{-2\phi}{*H_{\3}}\wedge H_{\3}
-\ft{1}{4}e^{-\phi}\big({*F_{\2}^I}\wedge F_{\2I}
+{*\widetilde{F}_{\2}^{\tilde{I}}}\wedge \widetilde{F}_{\2\tilde{I}}\big),\label{d4theory}
\eea
where the 2-form and 3-form field strengths are
\bea
F_{\2}^I&=&dA_{\1}^I+\ft{1}{2}g\epsilon^{IJK}
A_{\1J}\wedge A_{\1K}\,,\qquad
\widetilde{F}_{\2}^{\tilde{I}}=d\widetilde{A}_{\1}^{\tilde{I}}
+\ft{1}{2}g\epsilon^{\tilde{I}\tilde{J}\tilde{K}}
\widetilde{A}_{\1\tilde{J}}\wedge \widetilde{A}_{\1\tilde{K}}\,,\cr
H_{\3}&=&dB_{\2}+\ft{1}{4}\big(F_{\2}^I\wedge A_{\1I}+\widetilde{F}_{\2}^{\tilde{I}}\wedge \widetilde{A}_{\1{\tilde{I}}}\big)\cr
&&-\ft{1}{24}g\big(\epsilon_{IJ\gamma}A_{\1}^I\wedge A_{\1}^J\wedge A_{\1}^K+\epsilon_{\tilde{I}\tilde{J}\tilde{K}}
\widetilde{A}_{\1}^{\tilde{I}}\wedge \widetilde{A}_{\1}^{\tilde{J}}\wedge \widetilde{A}_{\1}^{\tilde{K}}\big)\,,\label{h3f2}
\eea
The 3-form field strength $H_\3$ satisfies the Bianchi identity
\be
dH_{\3}=\ft{1}{4}\big(F_{\2}^I\wedge F_{\2I}+\widetilde{F}_{\2}^{\tilde{I}}
\wedge \widetilde{F}_{\2\tilde{I}}\big)\,.\label{h3bianchi}
\ee
The complete set of equations of motion derived from the variations of the metric, $B_\2$, $(A_\1^I, \tilde A_1^{\tilde I})$ and the dilaton is given by
\bea
0&=&R_{\mu\nu}-\ft{1}{2}g_{\mu\nu}\mathcal{L}_4-\ft{1}{2}(\nabla_\mu\phi)(\nabla_\nu\phi)-
\ft{1}{4}e^{-2\phi}H_{\mu\nu}^2-\ft{1}{4}e^{-\phi}\Big(F_{\mu\rho}^I F_\nu^{I,\rho}+\widetilde{F}_{\mu\rho}^{\tilde{I}} \widetilde{F}_\nu^{\tilde{I},\rho}
\Big),\cr
0&=&\nabla_\mu\Big(\ft{1}{2}e^{-2\phi}H^{\mu\nu\rho}\Big),\cr
0&=&\nabla_\nu\Big(\ft{1}{4}e^{-2\phi}H^{\mu\nu\rho}\Big)A^I_\rho + \nabla_\nu\Big(\ft{1}{2}e^{-\phi}F^{I,\mu\nu}
\Big) + \ft{1}{4}e^{-2\phi}H^{\mu\nu\rho}F_{\nu\rho}^I- 
\ft12{g}e^{-\phi}\epsilon_{IJK}F^{J,\mu\nu}A_\nu^K\,,\cr
0&=&\nabla_\nu\Big(\ft{1}{4}e^{-2\phi}H^{\mu\nu\rho}\Big)\widetilde{A}^{\tilde{I}}_\rho
+ \nabla_\nu\Big(\ft{1}{2}e^{-\phi}\widetilde{F}^{\tilde{I},\mu\nu}
\Big) + \ft{1}{4}e^{-2\phi}H^{\mu\nu\rho}\widetilde{F}_{\nu\rho}^{\tilde{I}}
-\ft12{g}e^{-\phi}\epsilon_{\tilde{I}\tilde{J}\tilde{K}}
\widetilde{F}^{\tilde{J},\mu\nu}\widetilde{A}_\nu^{\tilde{K}}\,,\cr
0&=&\Box\phi+\ft{1}{6}e^{-2\phi}H^2_{\3}+\ft{1}{8}e^{-\phi}\Big(
F_{\mu\nu}^I F^{I,\mu\nu}+\widetilde{F}_{\mu\nu}^{\tilde{I}} \widetilde{F}^{\tilde{I},\mu\nu}\Big).
\eea
The fermionic sector consists of the gravitino $\psi_\mu$, the dilatino $\chi$, and a pair of gaugini $(\lambda^I, \tilde \lambda^{\tilde I})$. The Killing spinor equations arise from the vanishing of their supersymmetric transformation rules are
\bea
\delta \psi_a &=& \nabla_a \epsilon + \ft1{48} e^{-\phi} H_{bcd} [\Gamma^{bcd},\Gamma_a]\epsilon=0\,,\cr
\delta \chi &=&\ft12\nabla_a \phi \Gamma^a \epsilon+ \ft1{12} e^{-\phi} H_{abc} \Gamma^{abc} \epsilon=0\,,\cr
\delta\lambda^I&=&-\frac{1}{4\sqrt{2}}e^{-\frac{1}{2}\phi} F_{ab}^I\Gamma^{ab}  \epsilon=0\,,\cr 
\delta\tilde{\lambda}^{\tilde{I}}&=&-\frac{1}{4\sqrt{2}}e^{-\frac{1}{2}\phi} \widetilde{F}_{ab}^{\tilde{I}}\Gamma^{ab}  \epsilon =0\,.
\eea
Note that both $(I, \tilde I)=1,2,3 $ are the $SU(2)$ Yang-Mills indices. In this section, we construct a Euclidean NS(-1)-brane carrying the magnetic $H_\3$ flux, supported by the two copies of $SU(2)\times SU(2)$ 't Hooft Yang-Mills instantons, via the Bianchi identity \eqref{h3bianchi}.

\subsection{Single instanton configuration}

The NS(-1)-brane with a single point center can be written in spherically-symmetric coordinates. We find that the theory admits the following solution in the Einstein frame, 
\bea
ds_4^2&=&dr^2+\ft14 r^2(\bar{\sigma}_1^2+\bar{\sigma}_2^2+\bar{\sigma}_3^2)\,,\quad
A_{\1}^I = f\bar{\sigma}^i,\quad \widetilde{A}_{\1}^{\tilde{I}}=\tilde{f}\bar{\sigma}^{i},\quad \phi=\log H\,,\cr
H_{\3}&=& *dH\,,\qquad 
f=\frac{a^2}{r^2+ga^2},\qquad \tilde{f}=\frac{\tilde{a}^2}{r^2+g\tilde{a}^2}\,,\cr
H &=& 1+\frac{r^2+2ga^2}{g^2(r^2+ga^2)^2}+\frac{r^2+2g\tilde{a}^2}{g^2(r^2+g\tilde{a}^2)^2} \,,
\label{singlesol}
\eea
where $I=i=\tilde I$ and $\bar{\sigma}^i$ is the $SU(2)$ left-invariant 1-forms, given by
\be
\bar{\sigma}^1=\cos\psi d\theta+\sin\psi\sin\theta d\varphi,\quad \bar{\sigma}^2=-\sin\psi d\theta+\cos\psi\sin\theta d\varphi,\quad \bar{\sigma}^3=d\psi+\cos\theta d\varphi\,.
\ee
The metric for round unit $S^3$ is $d\bar \Omega_3^2 = \fft14(\bar{\sigma}_1^2+ \bar{\sigma}_2^2+\bar{\sigma}_3^2)$. The pair of the 't Hooft $SU(2)$ Yang-Mills instantons are both self-dual, i.e.~$F_{\2}^I=*F_{\2}^I$ and $\widetilde{F}_{\2}^{\tilde{I}}=*\widetilde{F}_{\2}^{\tilde{I}}$. The solution of $H_\3$ can also be expressed in terms of its 2-form potential $B_{\2}$:
\be
B_{\2}=-\frac{1}{2g^2}\cos\theta d\varphi\wedge d\psi\,.
\ee
To analyze the supersymmetry, it is natural to adopt the vielbein basis
\be
e^0 = dr\,,\qquad e^i=\frac{r}{2} \bar{\sigma}^i\,,\qquad i=1,2,3\,,
\ee
which can be used to express the form fields
\bea
H_{\3}&=&-H'e^1\wedge e^2\wedge e^3\,,\qquad F_{\2}^I=\frac{2f'}{r}e^0\wedge e^I+\frac{2f(gf-1)}{r^2}\epsilon^{IJK}e^J\wedge e^K\,,\cr
\widetilde{F}_{\2}^{\tilde{I}}&=&\frac{2\tilde{f}'}{r}e^0\wedge e^{\tilde{I}}
+\frac{2\tilde{f}(g\tilde{f}-1)}{r^2}\epsilon^{\tilde{I}\tilde{J}\tilde{K}}e^{\tilde{J}}\wedge e^{\tilde{K}}\,.
\eea
We find that the Killing spinor is given by
\be
\epsilon=H^{-\fft14} \epsilon_0\,,\qquad \Gamma^{0123}\epsilon=\epsilon\,.\label{d4ks}
\ee
Both the NS(-1)-brane and its Killing spinor for vanishing Yang-Mills instantons were obtained in \cite{Lu:2025nyo}. We observe that all the contributions from the Yang-Mills instantons enter the solution and its Killing spinor through the function $H$.

\subsection{Multi-instanton configurations}

To construct NS(-1)-brane solutions with multi-instantons, it is instructive to write e Euclidean 4-space in the Cartesian coordinates 
\be
ds_4^2=\delta_{\mu\nu}dx^\mu dx^\nu=dx_0^2+dx_1^2+dx_2^2+dx_3^2\,.\label{d4cart}
\ee
To accommodate the $SU(2)$ Yang-Mills fields, it is useful to introduce the 't Hooft tensors $\eta_{\mu\nu}^I$
\be
\eta^I=\frac{1}{2}\eta^I_{\mu\nu}dx^\mu\wedge dx^\nu=-dx^0\wedge dx^I+\frac{1}{2}\epsilon_{IJK}dx^J\wedge dx^K,
\ee
which means, $\eta^I_{0J}=-\delta^I_J=-\eta^I_{J0}$ and $\eta^I_{JK}=\epsilon_{IJK}$. The Ans\"atze for the Yang-Mills potentials are
\be
A_{\1}^I=-\eta^I_{\mu\nu}\partial^\mu fdx^\nu\,,\qquad \widetilde{A}_{\1}^{\tilde{I}}=-\eta^{\tilde{I}}_{\mu\nu}\partial^\mu \tilde{f}dx^\nu\,.
\ee
Imposing the self-duality condition requires that $(f, \tilde{f})$ be expressible in terms of harmonic functions $(f',\tilde{f}')$ on four-dimensional Euclidean space \eqref{d4cart}, namely
\be
f=-\frac{1}{g}\log f',\quad \tilde{f}=-\frac{1}{g}\log \tilde{f}'\,,\qquad\hbox{with}\qquad
\Box f'=0\,,\quad \Box \tilde{f}'=0\,.
\ee
To satisfy the equations of motion, the dilaton and the $H_\3$ field take the same form as in the previous single-center case discussed previously, except that here the function $H$ becomes more involved. We find the solutions are analogous to those in \cite{Lima:1999dn,Lima:1999xj}, given by
\bea
\phi&=&\log H,\qquad H_{\3}=*dH\,,\cr
H&=&1-\frac{1}{4g}\Box\big(f+\tilde{f}\big)+h,\qquad \Box h=0\,.\label{genHa}
\eea
It is useful to determine the 2-form potential $B_\2$ field. We find that $dB_\2 = {* dh}$. For any specific harmonic function $h$, we can obtain the explicit $B_\2$; however, a covariant expression for $B_\2$ corresponding to a generic harmonic function $h$ does not appear to exist.  The Killing spinor takes the same form as the single instanton case, given in \eqref{d4ks}, but now with the more involved function $H$ of \eqref{genHa}.

\subsection{Instanton moduli, topology change and NS(-1)-brane creation}

The four-dimensional NS(-1)-brane with all the Yang-Mills fields turned off was constructed in \cite{Lu:2025nyo}. In the string frame, the spherically symmetric metric takes the form
\be
ds_{\rm str}^2 = H (dr^2 + r^2 d\bar \Omega_3^2)\,,\qquad H = 1 + \fft{q}{r^2}\,.
\ee
The radial coordinate lies in the range  $r\in (0,\infty)$. The metric approaches the Euclidean $\mathbb R^4$ in the asymptotic $r\rightarrow \infty$. In the space middle $r\rightarrow 0$, the $S^3$ cone does not collapse to a point, but reaches to some finite radius $\sqrt{q}$ instead, and the radial line element becomes trivial, given by
\be
d\rho^2=\fft{dr^2}{r^2}\,,\qquad r=e^{\rho}\,.
\ee
Thus the topology of the space is $\mathbb R \times S^3$. As $\rho$ runs from $-\infty$ (corresponding to $r\rightarrow 0$) to $+\infty$ (corresponding to $r\rightarrow \infty$), the metric interpolates from the bolt geometry $\mathbb R \times S^3$ to asymptotic $\mathbb R^4$.  Since the string coupling is $g=\langle e^\phi\rangle$, the NS(-1)-brane bolt at the $r\rightarrow 0$ limit is infinitely strongly coupled. In general, since $H$ is a harmonic function of the Euclidean space, we can construct multi-center NS(-1)-branes
\be
ds_{\rm str}^2 = H^2 dy^i dy^i \,,\qquad H = 1 + \sum_{\alpha }^N \fft{q_\alpha}{|\vec y - \vec y_\alpha|^2}\,.
\ee
The metric approaches Euclidean $\mathbb R^4$ asymptotically as $y\rightarrow \infty$. At each center $\vec y_\alpha$, it reduces to the bolt geometry $\mathbb R\times S^3$.

The situation becomes more intriguing whence the $SU(2)\times SU(2)$ Yang-Mills fields are turned on. First we consider the single instanton situation, where the solution is spherically symmetric. The harmonic function $H$ in the previous case is now sourced by the 't Hooft instanton, as implied by the Bianchi identity \eqref{h3bianchi}. Consequently, the function $H$, given in \eqref{singlesol}, is no longer singular at $r=0$. In this case, the radial coordinate $r$ now ranges over $[0, \infty)$ instead of $(0,\infty)$. At $r=0$, the function $H$ reduces to some positive constant; therefore, $r=0$ corresponds to a smooth middle point of a Euclidean-signatured space, implying that the topology is $\mathbb R^4$. Interestingly, when either $a=0$ or $\tilde a=0$, the corresponding Yang-Mills fields vanish, the function $H$ does not reduce to 1, but instead gives rise to the NS(-1)-brane with $q=1/g$. The conserved 3-form charge is given by
\be
Q= \fft{1}{4\pi^2} \int dB_\2 = \fft{2}{g^2}\,.
\ee 
Thus, each $SU(2)$ instanton contributes one unit of $Q$ charge equal to $1/g^2$, independent of the instanton size $a$. Even when $a=0$ and the instanton disappears, the charge remains, a consequence of the topological nature of the Bianchi identity \eqref{h3bianchi}. Thus, the vanishing of the 't Hooft instanton results in the creation of the NS(-1)-brane, thereby changing the original metric topology $\mathbb R^4$  of the gauged solution to the bolt topology $\mathbb R\times S^3$ of the NS(-1)-brane.

The more general multi-center solution of multiple instantons is given by
\bea
H &=&  1 + \fft{1}{4g^2} \Box \Big[ \log \Big (1 + \sum_{\alpha=1}^N \fft{g\lambda_\alpha}{|\vec y - \vec y_\alpha|^2}\Big) +\log\Big(
 \sum_{\tilde \alpha=1}^{\tilde N} \fft{g \tilde \lambda_{\tilde \alpha}}{|\vec y - \vec y_{\tilde \alpha}|^2}\Big ) \Big]\nn\\
 &&+ \fft{1}{g^2}\sum_{\alpha=1}^N \fft{1}{|\vec y - \vec y_\alpha|^2} + \fft{1}{g^2}\sum_{\tilde \alpha=1}^{\tilde N} \fft{1}{|\vec y - \vec y_{\tilde \alpha}|^2} \,.
\eea
We observe that an NS(-1)-brane with bolt $\mathbb R\times S^3$  is created when a Yang-Mills instanton vanishes, or when two Yang-Mills instantons coalesce. However, throughout these changes in the instanton moduli, 
the 3-form magnetic flux remains conserved.

\section{Ten-dimensional string origins}

Having obtained the four-dimensional solution in the Euclidean signature of ${\cal N}=1$, $D=4$ supergravity, we now turn to its 10-dimensional string origins.  Focusing on the bosonic sector, the $D=4$ solution can be uplifted to the heterotic 5-brane of \cite{Strominger:1990et}, where the internal spacetime is the six-dimensional Minkowski worldvolume. The $SU(2)\times SU(2)$ Yang-Mills sector originate from the $E_8\times E_8$ or $SO(32)$ Yang-Mills fields of the heterotic string. However, this picture does not workout, when the supersymmetry is taken account. The reduction on the worldvoume Mink$_6$ necessarily preserves all supersymmetry, resulting in the maximum ${\cal N}=8$ supergravity. The supersymmetry reduction requires a Calabi-Yau manifold, which does not have an extension in Lorentzian signature. We have to Euclideanize the 5-brane world volume and replace it with the Calabi-Yau threefold.

Recently, an intriguing embedding of the ${\cal N}=1$, $D=4$ supergravity within the heterotic string was obtained, with a warped $\mathbb R\times T^{1,1}$ internal space. The $SU(2)\times SU(2)$ Yang-Mills fields are not direct descendants of the $E_8\times E_8$ or $SO(32)$ of the heterotic theory; rather, they originate from the isometry group of the $T^{1,1}$ space.  This consistent reduction ansatz was obtained by virtue of the recently proposed bosonic/heterotic duality, and the corresponding $S^3\times S^3$ reduction of the dual noncritical bosonic string theory is much simpler \cite{Ma:2025mvo,Lu:2025nyo}. In this new framework, the mysterious 't Hooft Yang-Mills instanton moduli become the much simpler geometric parameters.

\subsection{Bosonic string origin}

The low-energy effective Lagrangian of the $D=10$ noncritical string including the conformal anomaly term is
\bea
\check{\mathcal{L}}_{10}&=&\big(\check{R}-8g^2e^{\frac{1}{2}\check{\phi}}\big)\check{*}\oneone-
\frac{1}{2}\check{*}d\check{\phi}\wedge d\check{\phi}
-\frac{1}{2}e^{-\check{\phi}}\check{*}\check{H}_\3\wedge \check{H}_\3,\quad \check{H}_\3=d\check{B}_\2\,. \label{10D anomalous string theory}
\eea
The $S^3$ reduction of bosonic string is perhaps the simplest among the nontrivial Pauli-type sphere reductions in supergravities, and the reduction ansatz was presented in \cite{Chamseddine:1999uy,Cvetic:2000dm}. The generalization to $S^3\times S^3$ reduction on bosonic string with a cosmological constant was proposed in \cite{Ma:2025mvo}. It should be emphasized that the conformal anomaly term links the theory to the bosonic string, rather than any consistent truncation of a supergravity theory. Employing the procedure outlined in \cite{Ma:2025mvo}, we obtain the $D=10$ solution
\bea
d\check{s}_{10}^2&=&e^{\frac{3}{4}\phi}ds_{4}^2
+\frac{1}{4g^2}e^{-\fft14\phi}\left(h^I h_I+\tilde{h}^{\tilde{I}} \tilde{h}_{\tilde{I}}\right)\,,\cr
h^I&=&\sigma^I - g A_{\1}^I\,,\qquad \tilde h^{\tilde I} =\tilde{\sigma}^{\tilde I} - g \widetilde A_{\1}^{\tilde I}\,,\cr
\check{B}_{\2}&=&B_{\2}+\frac{1}{4g}(\cos\psi_1 A_{\1}^1-\sin\psi_1 A_{\1}^2)\wedge d\theta_1+\frac{1}{4g}A_{\1}^3\wedge d\psi_1\cr
&&+\frac{1}{4g}(\cos\psi_2 \widetilde{A}_{\1}^{\tilde{1}}-\sin\psi_2 \widetilde{A}_{\1}^{\tilde{2}})\wedge d\theta_2+\frac{1}{4g}\widetilde{A}_{\1}^{\tilde{3}}\wedge d\psi_2
\cr
&&-\frac{1}{4g^2}\left(\cos\theta_1 d\psi_1-g\mu_I A_{\1}^I\right)\wedge d\varphi_1 -\frac{1}{4g^2}\big(\cos\theta_2 d\psi_2-g\tilde{\mu}_{\tilde{I}} \widetilde{A}_{\1}^{\tilde{I}}\big)\wedge d\varphi_2\,.
\label{s3s3red}
\eea
Here, $I = 1, 2, 3$ and $\tilde{I} = \tilde{1}, \tilde{2}, \tilde{3}$ are the group indices associated with the two $SU(2)$ factors. To express the metrics of the two unit round $S^3$, we employ two sets of $SU(2)$ left-invariant 1-forms $\sigma^I$, $\tilde{\sigma}^{\tilde{I}}$
\bea
&&\sigma^1=\cos\psi_1 d\theta_1+\sin\psi_1\sin\theta_1 d\varphi_1\,,\qquad \tilde{\sigma}^{\tilde{1}}=\cos\psi_2 d\theta_2+\sin\psi_2\sin\theta_2 d\varphi_2\,,\cr
&&\sigma^2=-\sin\psi_1 d\theta_1+\cos\psi_1\sin\theta_1 d\varphi_1\,,\qquad \tilde{\sigma}^{\tilde{2}}=-\sin\psi_2 d\theta_2+\cos\psi_2\sin\theta_2 d\varphi_2\,,\cr
&&\sigma^3=d\psi_1+\cos\theta_1 d\varphi_1\,,\qquad \tilde{\sigma}^{\tilde{3}}=d\psi_2+\cos\theta_2 d\varphi_2\,.\label{sigma and the other sigma}
\eea
The coordinates $\{\theta_1, \varphi_1, \psi_1, \theta_2, \varphi_2, \psi_2\}$ represent two sets of Euler angles on $S^3 \times S^3$. The coordinate variables $\mu^I$ and $\tilde \mu^{\tilde I}$ will be explained in appendix \ref{app:formulae}. In the string frame, the metric is
\be
d\check{s}_{10}^2=H ds_{4}^2
+\frac{1}{4g^2}\left(h^I h_I+\tilde{h}^{\tilde{I}} \tilde{h}_{\tilde{I}}\right).
\ee
In the asymptotic $r\rightarrow \infty$ region, the metric approaches $\mathbb R^4\times S^3\times S^3 $, which is the Euclidean version of the Mink$_4\times S^3\times S^3$ vacuum discussed in \cite{Ma:2025mvo}. Near the interior region $r\rightarrow 0$, the metric remains regular regardless of whether the solution includes the Yang-Mills instanton sources. When the function $H$ is given by \eqref{singlesol}, the $r\rightarrow 0$ limit yields a geometry of the form $(S^3\times S^3)\ltimes \mathbb R^4$. In other words, it is an $S^3\times S^3$ bundle over $\mathbb R^4$. On the other hand, if the instanton sizes $a$ and/or $\tilde a$ shrink to zero and the Yang-Mills instantons disappear, the $r\rightarrow 0$ limit gives $\mathbb (S^3\times S^3)\times (S^3\times \mathbb R)$.  If the function $H$ includes contribution from both the gauge instanton and NS(-1)-brane charges, the topology becomes an $S^3\times S^3$ bundle over the bolt geometry $S^3\times \mathbb R$, i.e.~$\mathbb (S^3\times S^3)\ltimes (S^3\times \mathbb R)$. Thus, the instanton moduli becomes purely geometric; they are the parameters of the fibres in a space bundle.

It can be verified that the solution \eqref{s3s3red} admits Killing spinors, which satisfy the conditions for vanishing of the pseudo-supersymmetric transformations \cite{Lu:2011zx,Liu:2012jra,Lu:2011ku}
\bea
\delta \check \psi_M &=& \check D_M \check \epsilon - \fft1{96} e^{-\fft12\check \phi}\big(
\Gamma_M \Gamma^{PQR} - 12 \delta_M^P \Gamma^{QR}\big) \check H_{PQR}\, \check \epsilon -\fft{\rm i}{4\sqrt2} g\, e^{\fft14 \check\phi} \Gamma_M \check \epsilon =0\,,\nn\\
\delta \check \chi &=& -\fft12 \Gamma^M \partial_M\check{\phi}\,\check \epsilon -
\fft{1}{24} e^{-\fft12 \check \phi} \Gamma^{MNP} \check H_{MNP}\,\check \epsilon - \fft{\rm i}{\sqrt2} g\, e^{\fft14 \check \phi}\check \epsilon=0\,.\label{susy2}
\eea
The metric in \eqref{s3s3red} provides a natural basis for constructing the vielbein. However, to clarify how this solution maps to the corresponding one with $T^{1,1}$ internal space in the heterotic theory via the bosonic/heterotic duality, it is advantageous to perform a coordinate transformation 
\be
\varphi_1-\varphi_2=\chi_2\,,\qquad \varphi_1+\varphi_2=\chi_1\,,
\ee
and using several identities on $S^3$ provided in appendix \ref{app:formulae}, we can express the solution \eqref{sigma and the other sigma} as follows
\bea
d\check{s}_{10}^2&=&H^{\frac{3}{4}}\Big[dr^2+\frac{r^2}{4}(\bar{\sigma}_1^2+\bar{\sigma}_2^2+\bar{\sigma}_3^2)\Big]
+\frac{H^{-\frac{1}{4}}}{8g^2}\Bigg\{2\big(D\mu^I D\mu_I+D\tilde{\mu}^{\tilde{I}} D\tilde{\mu}_{\tilde{I}}\big)\cr
&&+\Big[
d\chi_1+\cos\theta_1d\psi_1-g\mu_IA^I_{\1}+\cos\theta_2d\psi_2-g\tilde{\mu}_{\tilde{I}}\widetilde{A}^{\tilde{I}}_{\1}\Big]^2\cr
&&+\Big[
d\chi_2+\cos\theta_1d\psi_1-g\mu_IA^I_{\1}-\cos\theta_2d\psi_2+g\tilde{\mu}_{\tilde{I}}\widetilde{A}^{\tilde{I}}_{\1}\Big]^2
\Bigg\}\,,\cr
\check{B}_{\2}&=&-\frac{1}{2g^2}\cos\theta d\varphi\wedge d\psi+\frac{1}{4g}(\cos\psi_1 A_{\1}^1-\sin\psi_1 A_{\1}^2)\wedge d\theta_1+\frac{1}{4g}A_{\1}^3\wedge d\psi_1\cr
&&+\frac{1}{4g}(\cos\psi_2 \widetilde{A}_{\1}^{\tilde{1}}-\sin\psi_2 \widetilde{A}_{\1}^{\tilde{2}})\wedge d\theta_2+\frac{1}{4g}\widetilde{A}_{\1}^{\tilde{3}}\wedge d\psi_2
\cr
&&-\frac{1}{8g^2}\big(\cos\theta_1 d\psi_1-g\mu_I A_{\1}^I+\cos\theta_2 d\psi_2-g\tilde{\mu}_{\tilde{I}} \widetilde{A}_{\1}^{\tilde{I}}\big)\wedge d\chi_1 \cr
&&-\frac{1}{8g^2}\big(\cos\theta_1 d\psi_1-g\mu_I A_{\1}^I-\cos\theta_2 d\psi_2+g\tilde{\mu}_{\tilde{I}} \widetilde{A}_{\1}^{\tilde{I}}\big)\wedge d\chi_2\,,\cr
\check{\phi}&=&\frac{1}{2}\log H\,.\label{bosonic vacuum}
\eea

To evaluate the Killing spinor equations \eqref{susy2}, we express both the metric and the $H_{\3}$ field in terms of vielbeins. Using the identities provided in appendix \ref{app:formulae}, we have
\bea
D\mu^I D\mu_I=\varepsilon^a\varepsilon^a\,,\qquad
D\tilde{\mu}^{\tilde{I}} D\tilde{\mu}_{\tilde{I}}=\tilde{\varepsilon}^{\tilde{a}}
\tilde{\varepsilon}^{\tilde{a}}\,.
\eea
The metric can be rewritten as
\bea
d\check{s}_{10}^2&=&H^{\frac{3}{4}}\Big[dr^2+\frac{r^2}{4}(\bar{\sigma}_1^2+\bar{\sigma}_2^2+\bar{\sigma}_3^2)\Big]
+\frac{H_a^{-\frac{1}{4}}}{8g^2}\Big[2\big(\varepsilon^a\varepsilon^a+\tilde{\varepsilon}^{\tilde{a}}
\tilde{\varepsilon}^{\tilde{a}}\big)+\Sigma_1^2+\Sigma_2^2
\Big]\,,\cr
\Sigma_1&=&d\chi_1+\cos\theta_1d\psi_1-g\mu_IA^I_{\1}+\cos\theta_2d\psi_2-g\tilde{\mu}_{\tilde{I}}\widetilde{A}^{\tilde{I}}_{\1}\,,\cr
\Sigma_2&=&d\chi_2+\cos\theta_1d\psi_1-g\mu_IA^I_{\1}-\cos\theta_2d\psi_2+g\tilde{\mu}_{\tilde{I}}\widetilde{A}^{\tilde{I}}_{\1}\,,\label{bosonic metric vielbein}
\eea
and now the natural vielbein choice is
\bea
&&\check e^0 = H^{\fft38} dr\,,\qquad \check e^I = \ft12 r H^{\fft38} \bar{\sigma}^I\,,\qquad \check e^{\tilde{I}} = \ft12 r H^{\fft38} \bar{\sigma}^{\tilde{I}}\,,\cr
&&\check e^a= \fft{H^{-\fft18}}{2g}\,\varepsilon^a\,,\qquad
\check e^{\tilde{a}}=\fft{H^{-\fft18}}{2g} \,\tilde{\varepsilon}^{\tilde{a}}\,,\qquad
\check e^8=\fft{H^{-\fft18}}{2\sqrt2\,g} \Sigma_1\,,\qquad
\check e^9 = \fft{H^{-\fft18}}{2\sqrt2\,g} \Sigma_2\,.\label{d10bstrviel2}
\eea
The form fields under this vielbein choice are
\bea
\check{H}_{\3}&=&-H'H^{-\fft98}\check{e}^1\wedge \check{e}^2\wedge \check{e}^3+\frac{H^{\fft18}}{2}K_I^a F_{\2}^I\wedge \check{e}^a
+\frac{H^{\fft18}}{2}\widetilde{K}_{\tilde{I}}^{\tilde{a}}
\widetilde{F}_{\2}^{\tilde{I}}\wedge \check{e}^{\tilde{a}}\cr
&&+\frac{H^{\fft18}}{2\sqrt{2}}\Big[\mu_I F_{\2}^I+\tilde{\mu}_{\tilde{I}} \widetilde{F}_{\2}^{\tilde{I}}+2gH^{\fft14}\big(\epsilon_{ab}\, \check{e}^a\wedge \check{e}^b+\epsilon_{\tilde{a}\tilde{b}}\, \check{e}^{\tilde a}\wedge \check{e}^{\tilde b}\big)\Big] \wedge\check{e}^8\cr
&&+\frac{H^{\fft18}}{2\sqrt{2}}\Big[\mu_I F_{\2}^I-\tilde{\mu}_{\tilde{I}} \widetilde{F}_{\2}^{\tilde{I}}+2gH^{\fft14}\big(\epsilon_{ab}\, \check{e}^a\wedge \check{e}^b-\epsilon_{\tilde{a}\tilde{b}}\, \check{e}^{\tilde a}\wedge \check{e}^{\tilde b}\big)\Big] \wedge\check{e}^9\,,\cr
F_{\2}^I&=&H^{-\fft34}\Big[\frac{2f'}{r}\check{e}^0\wedge \check{e}^I+\frac{2f(gf-1)}{r^2}\epsilon^{IJK}\check{e}^J\wedge \check{e}^K\Big],\cr
\widetilde{F}_{\2}^{\tilde{I}}&=&H^{-\fft34}\Big[\frac{2\tilde{f}'}{r}\check{e}^0\wedge \check{e}^{\tilde{I}}
+\frac{2\tilde{f}(g\tilde{f}-1)}{r^2}\epsilon^{\tilde{I}\tilde{J}\tilde{K}}\check{e}^{\tilde{J}}\wedge \check{e}^{\tilde{K}}\Big].
\eea
We find that the Killing spinors are 
\be
\check \epsilon =H^{-\fft1{16}} \Big( e^{\fft12{\rm i} \chi_1} \check \epsilon_{1,+} +e^{-\fft12{\rm i} \chi_1} \check \epsilon_{1,-}+  e^{\fft12{\rm i} \chi_2} \check \epsilon_{2,+} + e^{-\fft12{\rm i} \chi_2} \check \epsilon_{2,-} \Big),
\ee
where the constant spinors $\check \epsilon_{i,\pm}$ satisfy the projections 
\bea
\Gamma^{0123} \check \epsilon_{i,\pm} = \check\epsilon_{i,\pm}\,,\qquad i=1,2\,,\cr
\Gamma^{45}\, \check \epsilon_{1,\pm} = \Gamma^{67}\, \check \epsilon_{1,\pm} = -
{\rm i}\, \Gamma^8\, \check \epsilon_{1,\pm} &=& \mp {\rm i}\, \check \epsilon_{1,\pm}\,,\nn\\
\Gamma^{45}\, \check \epsilon_{2,\pm} = -\Gamma^{67}\, \check \epsilon_{2,\pm} = -
{\rm i}\, \Gamma^9\, \check \epsilon_{2,\pm} &=& \mp {\rm i}\, \check \epsilon_{2,\pm}\,.
\eea
The Killing spinor takes the same form as that in \cite{Lu:2025nyo}, where the Yang-Mills fields were turned off. All the Yang-Mills contribution to the Killing spinors enters solely through the function $H$. As was discussed in \cite{Lu:2025nyo}, $\chi_2$ plays a role of bridging coordinate in the bosonic/heterotic duality. Consequently , the Killing spinors $\check \epsilon_{2,\pm}$ will not survive the duality map. In contrast, the Killing spinors $\check\epsilon_{1,\pm}$ will survive the map and become the heterotic Killing spinors after the duality transformation.

\subsection{Heterotic string origin}

The relevant minimum bosonic field content consists of the metric, dilaton scalar $\hat\phi$, 2-form potential $\hat B_\2$ and one matter $U(1)$ vector $\hat A_\1$. The Lagrangian is
\bea
\hat{\mathcal{L}}_{10}&=& \hat{R}\, {\hat *\oneone}-\ft12 {\hat * d\hat \phi}\wedge d\hat \phi -\ft12 e^{-\hat\phi} {\hat *\hat H_\3} \wedge \hat H_\3 -\ft{1}{2 }e^{-\frac{1}{2}\hat{\phi}}{\hat *\hat F_\2}\wedge \hat F_\2\,,\cr
\hat H_\3&=&dB_\2+\frac{1}{2}\hat F_\2\wedge\hat A_\1.\label{heterotic}
\eea
The inclusion of $\hat F_\2$ implies that the theory cannot be embedded into either type IIA or IIB maximal supergravities. Following the reduction ansatz of \cite{Ma:2025mvo}, we find that the instanton/wormhole solution of $D=4$ becomes
\bea
d\hat{s}_{10}^2&=&e^{2mu}\Bigg\{H^{\frac{3}{4}}\Big[dr^2+\frac{r^2}{4}(\bar{\sigma}_1^2+\bar{\sigma}_2^2+\bar{\sigma}_3^2)\Big]
+H^{-\frac{1}{4}}\Big[\frac{1}{4g^2}\big(D\mu^I D\mu_I+D\tilde{\mu}^{\tilde{I}} D\tilde{\mu}_{\tilde{I}}\big)\cr
&&+\frac{1}{8g^2}\big(
d\chi_1+\cos\theta_1d\psi_1-g\mu_IA^I_{\1}+\cos\theta_2d\psi_2-g\tilde{\mu}_{\tilde{I}}\widetilde{A}^{\tilde{I}}_{\1}\big)^2+du^2
\Big]\Bigg\}\,,\cr
\hat{B}_{\2}&=&-\frac{1}{2g^2}\cos\theta d\varphi\wedge d\psi+\frac{1}{4g}(\cos\psi_1 A_{\1}^1-\sin\psi_1 A_{\1}^2)\wedge d\theta_1+\frac{1}{4g}A_{\1}^3\wedge d\psi_1\cr
&&+\frac{1}{4g}(\cos\psi_2 \widetilde{A}_{\1}^{\tilde{1}}-\sin\psi_2 \widetilde{A}_{\1}^{\tilde{2}})\wedge d\theta_2+\frac{1}{4g}\widetilde{A}_{\1}^{\tilde{3}}\wedge d\psi_2
\cr
&&-\frac{1}{8g^2}\big(\cos\theta_1 d\psi_1-g\mu_I A_{\1}^I+\cos\theta_2 d\psi_2-g\tilde{\mu}_{\tilde{I}} \widetilde{A}_{\1}^{\tilde{I}}\big)\wedge d\chi_1 \,,\cr
\hat{A}_{\1}&=&-\frac{1}{2g}\big(\cos\theta_1 d\psi_1-g\mu_I A_{\1}^I-\cos\theta_2 d\psi_2+g\tilde{\mu}_{\tilde{I}} \widetilde{A}_{\1}^{\tilde{I}}\big)\,,\cr
\hat{\phi}&=&\frac{1}{2}\log H-4mu\,.\label{Heterotic vacuum}
\eea
In the string frame, the metric is
\bea
d\hat{s}_{10}^2&=&H\big(dr^2+\ft14 r^2(\bar{\sigma}_1^2+\bar{\sigma}_2^2+\bar{\sigma}_3^2)\big)
+\frac{1}{4g^2}\big(D\mu^I D\mu_I+D\tilde{\mu}^{\tilde{I}} D\tilde{\mu}_{\tilde{I}}\big)\cr
&&+\frac{1}{8g^2}\big(
d\chi_1+\cos\theta_1d\psi_1-g\mu_IA^I_{\1}+\cos\theta_2d\psi_2-g\tilde{\mu}_{\tilde{I}}
\widetilde{A}^{\tilde{I}}_{\1}\big)^2+du^2\,.
\eea
The metric is smooth, with the radial coordinate running from $r=0$ to asymptotic $r\rightarrow \infty$, where the geometry approaches $\mathbb R^5\times T^{1,1}$. Depending on the instanton moduli, the bolt topology can be either a direct product of $T^{1,1}\times S^3 \times \mathbb R^2$, or an $\mathbb R\times T^{1,1}$ bundle over $\mathbb R^4$, or an $\mathbb R\times T^{1,1}$ bundle over $\mathbb R\times S^3$. Again, the instanton moduli now serve as geometric parameters in the $T^{1,1}$ fibre space.

To analyze the supersymmetry, following the approach used in the anomalous bosonic string case \eqref{bosonic metric vielbein}, we also rewrite the metric
\crampest{\bea
d\hat{s}_{10}^2&=&e^{2mu}\bigg\{H^{\frac{3}{4}}\Big[dr^2+\frac{r^2}{4}(\bar{\sigma}_1^2+\bar{\sigma}_2^2+\bar{\sigma}_3^2)\Big]
+H^{-\frac{1}{4}}\Big[\frac{1}{32m^2}\big(\varepsilon^a\varepsilon^a+\tilde{\varepsilon}^{\tilde{a}}
\tilde{\varepsilon}^{\tilde{a}}\big)+\frac{1}{64m^2}\Sigma_1^2+du^2
\Big]\bigg\},
\eea}
and choose the vielbein
\bea
&&\hat e^0 = e^{mu}H^{\fft38} dr\,,\qquad \hat e^I = \frac{r}{2}e^{mu}H^{\fft38}\bar{\sigma}^I\,,\qquad \hat e^{\tilde{I}} = \frac{r}{2}e^{mu}H^{\fft38}\bar{\sigma}^{\tilde{I}}\,,\cr
&&\hat e^a= \fft{e^{mu}H^{-\fft18}}{4\sqrt{2}m}\,\varepsilon^a\,,\quad
\hat e^{\tilde{a}}=\fft{e^{mu}H^{-\fft18}}{4\sqrt{2}m} \,\tilde{\varepsilon}^{\tilde{a}}\,,\quad
\hat e^8=\fft{e^{mu}H^{-\fft18}}{8m} \Sigma_1\,,\quad
\hat e^9 = e^{mu}H^{-\fft18}du\,.\label{Heterotic d10bstrviel2}
\eea
The form fields under these vielbein basis are
\bea
\hat{H}_{\3}&=&-e^{-3mu}H'H^{-\fft98}\hat{e}^1\wedge \hat{e}^2\wedge \hat{e}^3+\frac{e^{-mu}H^{\fft18}}{2}K_I^a F_{\2}^I\wedge \hat{e}^a
+\frac{e^{-mu}H^{\fft18}}{2}\widetilde{K}_{\tilde{I}}^{\tilde{a}}
\widetilde{F}_{\2}^{\tilde{I}}\wedge \hat{e}^{\tilde{a}}\cr
&&+\frac{e^{-mu}H^{\fft18}}{2\sqrt{2}}\Big[\mu_I F_{\2}^I+\tilde{\mu}_{\tilde{I}} \widetilde{F}_{\2}^{\tilde{I}}+4\sqrt{2}me^{-2mu}H^{\fft14}\big(\epsilon_{ab}\, \hat{e}^a\wedge \hat{e}^b+\epsilon_{\tilde{a}\tilde{b}}\, \hat{e}^{\tilde a}\wedge \hat{e}^{\tilde b}\big)\Big] \wedge\hat{e}^8\,,\cr
\hat{F}_{\2}&=&\ft{1}{2}(\mu_\alpha F_{\2}^\alpha-\tilde{\mu}_{\tilde{\alpha}} F_{\2}^{\tilde{\alpha}})
+2\sqrt{2}me^{-2mu}H^{\fft14}\Big( \epsilon_{ab}\, \hat{e}^a\wedge \hat{e}^b
-\epsilon_{\tilde{a}\tilde{b}}\, \hat{e}^{\tilde a}\wedge \hat{e}^{\tilde b}\Big)\,,\cr
F_{\2}^I&=&e^{-2mu}H^{-\fft34}\Big[\frac{2f'}{r}\hat{e}^0\wedge \hat{e}^I+\frac{2f(gf-1)}{r^2}\epsilon^{IJK}\hat{e}^J\wedge \hat{e}^K\Big]\,,\cr
\widetilde{F}_{\2}^{\tilde{I}}&=&e^{-2mu}H^{-\fft34}\Big[\frac{2\tilde{f}'}{r}\hat{e}^0\wedge \hat{e}^{\tilde{I}}
+\frac{2\tilde{f}(g\tilde{f}-1)}{r^2}\epsilon^{\tilde{I}\tilde{J}\tilde{K}}\hat{e}^{\tilde{J}}\wedge \hat{e}^{\tilde{K}}\Big]\,.
\eea
Once again, we find that the solution is preserves supersymmetry and admits Killing spinors given by
\be
\hat{\epsilon}=e^{\frac{mu}{2}} {H^{-\frac{1}{16}}}\Big(e^{\frac{\mathrm{i}\chi_1}{2}}\hat{\epsilon}_{0,+}+
e^{-\frac{\mathrm{i}\chi_1}{2}}\hat{\epsilon}_{0,-}
\Big)\,,
\ee
where the constant spinors $\hat \epsilon_{0,\pm}$ satisfy the projections 
\bea
&&\Gamma^{45}\hat \epsilon_{0,\pm} = \Gamma^{67}\hat \epsilon_{0,\pm}=\Gamma^{89} \hat \epsilon_{0,\pm}
=\mp {\rm i}\, \hat \epsilon_{0,\pm}\,,\cr
&&\Gamma^{0123} \hat\epsilon_{0,\pm} = \hat{\epsilon}_{0,\pm}\,.
\eea
It follows from the above, the $\Gamma^{11}$ projection of the Killing spinors is
\be
\Gamma^{11} \hat\epsilon_{0,\pm} =\pm {\rm i}\, \hat\epsilon_{0,\pm}\,.
\ee
Thus, a specific choice of chirality, which we must impose in the heterotic theory,  further projects out half the Killing spinors, either $\hat \epsilon_{0,+}$ or $\hat \epsilon_{0,-}$, depending on the convention.

\section{New BPS solution in the SS model}

As a brief digression, we note that when the ${\cal N}=1$, $D=4$ theory involves only a single copy of the $SU(2)$ Yang-Mills field, it can also be obtained via a Pauli-type $S^2$ reduction \cite{Gibbons:2003gp} of the $D=6$ SS model \cite{Salam:1984cj}. The bosonic sector Lagrangian of the latter is
\bea
\bar{\mathcal{L}}_{6}&=&
\big(\bar{R} -4g^2e^{\fft1{\sqrt2}\bar{\phi}}\big){\bar *\oneone}
-\ft12 {\bar *d\bar \phi}\wedge d\bar\phi -\ft12 e^{-\sqrt2 \bar{\phi}}
{\bar * \bar H}_\3\wedge \bar H_\3
-\ft12 e^{-\fft{1}{\sqrt2}\bar{\phi}} {\bar *\bar F}_\2\wedge \bar F_\2\,,\cr
\bar{H}_{\3}&=&d\bar{B}_{\2}+\ft{1}{2}\bar{F}_{\2}\wedge \bar{A}_{\1},\qquad \bar{F}_{\2}=d\bar{A}_{\1}\,. 
\eea
(Note that we need to rescale the gauge coupling $g$ by $g\rightarrow \sqrt2\, g$ to match the convention in \cite{Gibbons:2003gp}.) The Killing spinor equations corresponding to the vanishing of the supersymmetric transformation of the fermions are
\bea
&&\big(e^{-\frac{\sqrt{2}}{4}\bar{\phi}}\bar{F}_{\mu\nu}\Gamma^{\mu\nu}
-4\sqrt{2}\mathrm{i}ge^{\frac{\sqrt{2}}{4}\bar{\phi}}\big)\bar{\epsilon}=0\,,\cr
&&\big(-\sqrt{2}\Gamma^\mu\partial_\mu\bar{\phi}-
\frac{1}{6}e^{-\frac{\bar{\phi}}{\sqrt{2}}}\bar{H}_{\mu\nu\rho}^{-}\Gamma^{\mu\nu\rho}
\big)\bar{\epsilon}=0\,,\cr
&&\big(\bar{\nabla}_\mu-\mathrm{i}\frac{g}{\sqrt{2}}\bar{A}_\mu
+\frac{1}{48}e^{-\frac{\bar{\phi}}{\sqrt{2}}}
\bar{H}_{\alpha\beta\gamma}^{+}\Gamma^{\alpha\beta\gamma}\Gamma_\mu
\big)\bar{\epsilon}=0\,.
\eea
The six-dimensional theory is chiral with ${\cal N}=(1,0)$ supersymmetry, and its Killing spinor equations differ significantly from the previous examples. In particular, in Euclidean signature, $\bar{H}_{\3}^{\pm}$ can only be complex self-dual or anti-self-dual. They are given by
\be
\bar{H}_{\mu\nu\rho}^{\pm}=\frac{1}{2}(\bar{H}_{\mu\nu\rho}\pm \mathrm{i}\bar{*}\bar{H}_{\mu\nu\rho})\,.
\ee
By consistently switching off $\widetilde{A}_{\1}^{\tilde{I}}$ in equation (12) (i.e., setting $\tilde{a} = 0$), we obtain a 4D supergravity theory coupled to a single $SU(2)$ Yang-Mills field. It can be consistently uplifted through an $S^2$ reduction \cite{Gibbons:2003gp,Ma:2025mvo}
\bea
d\bar{s}_{6}^2&=&
e^{\frac{\phi}{2}}ds_{4}^2+\frac{1}{4g^2}e^{-\frac{\phi}{2}} D\mu^\alpha D\mu_\alpha,\cr
D\mu^\alpha D\mu_\alpha&=&(d\theta_1-gA^1_{\1}\cos\psi_1+gA^2_{\1}\sin\psi_1)^2\cr
&&+\sin^2\theta_1(d\psi_1+gA^1_{\1}\cot\theta_1\sin\psi_1+gA^2_{\1}
\cot\theta_1\cos\psi_1-gA^3_{\1})^2\,,\cr
\bar{A}_{\1}&=&-\frac{1}{\sqrt{2}g}\cos\theta_1 d\psi_1+\frac{1}{\sqrt{2}}\mu_\alpha A_{\1}^\alpha\,,\cr
\bar{B}_{\2}&=&B_{\2}+\frac{1}{4g}(\cos\psi_1 A_{\1}^1-\sin\psi_1 A_{\1}^2)\wedge d\theta_1+\frac{1}{4g}A_{\1}^3\wedge d\psi_1\,,\cr
\bar{\phi}&=&\frac{\phi}{\sqrt{2}}=\frac{1}{\sqrt2} \log H.\label{s2red}
\eea
In the string frame, the metric interpolates $\mathbb R_4\times S^2$ of the asymptotic $r\rightarrow \infty$ and $S^2\ltimes \mathbb R^4$ with the Yang-Mills turned on, or $S^2\ltimes S^3\times \mathbb R^1$ with the Yang-Mills turned off.

To study the supersymmetry, we choose the follow natural choice of the vielbein
\bea
&&\bar e^0 = H^{\fft14} dr\,,\qquad \bar e^I = \ft12 r H^{\fft14} \bar{\sigma}^I\,,\qquad
\bar e^4= \fft{H^{-\fft14}}{2g}\,(d\theta_1-gA^1_{\1}\cos\psi_1+gA^2_{\1}\sin\psi_1)\,,\cr
&&\bar e^{5}=\fft{H^{-\fft14}}{2g} \,\sin\theta_1(d\psi_1+gA^1_{\1}\cot\theta_1\sin\psi_1+gA^2_{\1}
\cot\theta_1\cos\psi_1-gA^3_{\1})\,.
\eea
Although the Killing spinor equations in $D=6$ differ significantly from the previous cases, we find that the solution still admits Killing spinors, given by
\be
\bar{\epsilon}=H^{-\frac{1}{8}}\big(\bar{\epsilon}_++\bar{\epsilon}_-\big)\,,
\ee
with the projection
\be
\Gamma^{0123}\bar{\epsilon}_\pm=\bar{\epsilon}_\pm=\pm \mathrm{i}\,\Gamma^{01}\bar{\epsilon}_\pm
=\mp \mathrm{i}\,\Gamma^{23}\bar{\epsilon}_\pm=-\Gamma^7\bar{\epsilon}_\pm.
\ee

\section{Conclusions}

The purpose of this work is twofold. One is to construct the BPS NS(-1)-brane with $SU(2)\times SU(2)$ 't Hooft instanton sources in ${\cal N}=1$, $D=4$ supergravity. By uplifting to the heterotic string compactified on internal warped $\mathbb R\times T^{1,1}$ space, we obtain a new class of BPS solutions in the heterotic theory. We examine how the gauge instanton sources affect the string metrics. While the magnetic string flux is preserved under the variations of the instanton moduli, the topology of the string metrics can change, leading to $\mathbb R^4$ with source, or the $\mathbb R\times S^3$ bolt topology when the instanton size shrinks to zero. Gravitational instantons \cite{Gibbons:1978tef} with Killing spinors in four dimensions typically involves the NUT-type $\mathbb R^4$ topology or $\mathbb R^2\times S^2$ topology \cite{Hawking:1976jb}. Our new solutions provide new topological examples within the class of gravitational instantons. The geometric origin of the $SU(2)\times SU(2)$ Yang-Mills implies a rich fibre structure of the internal $T^{1,1}$ space. The phenomenon of brane creation via the mysterious 't Hooft instanton moduli can now be simply explained as a manifestation of the parameter space of pure geometry.

The other purpose of this work is to test the recently proposed bosonic/heterotic duality, which relates the a class of heterotic theories with linear dilaton along $\mathbb R$ to noncritical string with the conformal anomaly term at the T-duality fixed point. We explicitly construct the dual pair in the bosonic string theory and we find that the solution also admits Killing spinors, enabling to understand how the Killing spinors were related in this duality. This exercise provides further support to the proposed duality. Relatedly, we obtain new BPS solutions in the SS model, based on the technique developed in this paper. The Killing spinor equations of the $D=6$ SS chiral model differs significantly from those in other theories, yet Killing spinors exist in this class of solutions. This suggests a web of connections among the heterotic string, bosonic string and the SS model.

\section*{Acknowledgement}

L.M.~is supported in part by National Natural Science Foundation of China (NSFC) grant No.~12447138, Postdoctoral Fellowship Program of CPSF Grant No.~GZC20241211, the China Postdoctoral Science Foundation under Grant No.~2024M762338 and the National Key Research and Development Program No.~2022YFE0134300. H.L.~is supported in part by the National Natural Science Foundation of China (NSFC) grants No.~12375052 and No.~11935009. The work is also supported in part by the Tianjin University Self-Innovation Fund Extreme Basic Research Project Grant No.~2025XJ21-0007.

\section*{Appendix}

\appendix

\section{Useful formulae}
\label{app:formulae}

In this appendix, we present some useful formulae associated with the 2-sphere, 3-sphere and their relations. The material was largely presented in \cite{Gibbons:2003gp}. We list here for self consistency of the work. Consider a unit round $S^2$ parameterized by $\mu^I \mu^I =1$, with
\be
\mu^1 = \sin\theta \sin\psi\,,\qquad
\mu^2 = \sin\theta \cos\psi\,,\qquad
\mu^3 = \cos\theta\,.
\ee
Its metric is given by
\be
ds_2^2 = d\theta^2 + \sin^2\theta d\psi^2= d\mu^I d\mu^I = e^a e^a\,,
\ee
The three Killing vectors are given by
\be
K_I^m =\epsilon^{mn} \partial_n \mu_I\,.
\ee
Thus we have
\bea
K_1 = \cos\psi\, \fft{\partial}{\partial\theta} - \sin\psi \cot\theta\, \fft{\partial}{\partial\psi}\,,\quad
K_2 = - \sin\psi\, \fft{\partial}{\partial\theta} - \cos\psi \cot\theta\,
\fft{\partial}{\partial\psi}\,,\quad
K_3 = \fft{\partial}{\partial\psi}\,.
\eea
We have various useful identities
\be
K_I^m K_I^n = g^{mn}\,,\qquad g_{mn} K_I^m K_J^n = \delta_{IJ} - \mu_I \mu_J\,,\qquad
K_I^a e^a = -\epsilon_{IJK} \mu^J d\mu^K\,.
\ee
\be
K_I = -\epsilon_{IJK} \mu^J \wedge d\mu^K\,,\qquad
dK_I = 2\mu^I \Omega_\2 = - \epsilon_{IJK} d\mu^J
\wedge d\mu^K\,.
\ee
Note that $S^3$ can be viewed as a $U(1)$ bundle of $S^2$.  We can thus rewrite
$h^I$, defined by \eqref{s3s3red}, by
\bea
h^I&=&-\epsilon^{IJK}\mu_J D\mu_K+\mu^I\sigma\,,\qquad D\mu^I=d\mu^I+g\epsilon^{IJK}A_{\1J}\mu_K\,,\cr
\sigma &=&  d\varphi+\mathcal{A}_{\1}\,,\qquad \mathcal{A}_{\1}=\cos\theta d\psi-g\mu_I A_{\1}^I\,.
\eea
The metric $h_I h^I$ can be expressed as Kaluza-Klein form
\bea
h_I h^I&=&D\mu^I D\mu_I+\sigma^2,\cr
D\mu^I D\mu_I&=&(d\theta-gA^1_{\1}\cos\psi+gA^2_{\1}\sin\psi)^2\cr
&&+\sin^2\theta(d\psi+gA^1_{\1}\cot\theta\sin\psi+gA^2_{\1}
\cot\theta\cos\psi-gA^3_{\1})^2=\varepsilon^a \varepsilon^a.
\eea
Here we have
\be
\varepsilon^a = e^a - g K^a_I A^I_\1\,.\label{varep}
\ee
An important identity for form fields is that
\be
\ft12 \epsilon_{IJK} \mu^I D\mu^J\wedge D\mu^K = -\ft12 \epsilon_{ab}\, \varepsilon^a\wedge\varepsilon^b\,.
\ee
These formulae are also valid for the $S^3$ with tilded coordinates.


\begin{thebibliography}{99}

\bibitem{Goto:1971ce}
Y.~Nambu, “Duality and hydrodynamics,” Lectures at the Copenhagen conference,
 1970; T.~Goto,
``Relativistic quantum mechanics of one-dimensional mechanical continuum and subsidiary condition of dual resonance model,''
Prog. Theor. Phys. \textbf{46}, 1560-1569 (1971)
doi:10.1143/PTP.46.1560

\bibitem{Polyakov:1981rd}
A.M.~Polyakov,
``Quantum geometry of bosonic strings,''
Phys. Lett. B \textbf{103}, 207-210 (1981)
doi:10.1016/0370-2693(81)90743-7


\bibitem{Dabholkar:1990yf}
A.~Dabholkar, G.W.~Gibbons, J.A.~Harvey and F.~Ruiz Ruiz,
``Superstrings and solitons,''
Nucl. Phys. B \textbf{340}, 33-55 (1990)
doi:10.1016/0550-3213(90)90157-9

\bibitem{Duff:1994an}
M.J.~Duff, R.R.~Khuri and J.X.~Lu,
``String solitons,''
Phys. Rept. \textbf{259}, 213-326 (1995)
doi:10.1016/0370-1573(95)00002-X
[arXiv:hep-th/9412184 [hep-th]].

\bibitem{Gueven:1992hh}
R.~Gueven,
Phys. Lett. B \textbf{276}, 49-55 (1992)
doi:10.1201/9781482268737-16

\bibitem{Horowitz:1991cd}
G.T.~Horowitz and A.~Strominger,
``Black strings and $p$-branes,''
Nucl. Phys. B \textbf{360}, 197-209 (1991)
doi:10.1016/0550-3213(91)90440-9

\bibitem{Duff:1991pea}
M.J.~Duff and J.X.~Lu,
``The selfdual type IIB superthreebrane,''
Phys. Lett. B \textbf{273}, 409-414 (1991)
doi:10.1016/0370-2693(91)90290-7

\bibitem{Gibbons:1994vm}
G.W.~Gibbons, G.T.~Horowitz and P.K.~Townsend,
``Higher dimensional resolution of dilatonic black hole singularities,''
Class. Quant. Grav. \textbf{12}, 297-318 (1995)
doi:10.1088/0264-9381/12/2/004
[arXiv:hep-th/9410073 [hep-th]].

\bibitem{Strominger:1990et}
A.~Strominger,
``Heterotic solitons,''
Nucl. Phys. B \textbf{343}, 167-184 (1990)
[erratum: Nucl. Phys. B \textbf{353}, 565-565 (1991)]
doi:10.1016/0550-3213(90)90599-9

\bibitem{Duff:1996cf}
M.J.~Duff, H.~L\"u and C.N.~Pope,
``Heterotic phase transitions and singularities of the gauge dyonic string,''
Phys. Lett. B \textbf{378}, 101-106 (1996)
doi:10.1016/0370-2693(96)00420-0
[arXiv:hep-th/9603037 [hep-th]].

\bibitem{Cvetic:2000mh}
M.~Cveti\v c, H.~L\"u and C.N.~Pope,
``Brane resolution through transgression,''
Nucl. Phys. B \textbf{600}, 103-132 (2001)
doi:10.1016/S0550-3213(01)00050-5
[arXiv:hep-th/0011023 [hep-th]].

\bibitem{Klebanov:2000hb}
I.R.~Klebanov and M.J.~Strassler,
``Supergravity and a confining gauge theory: Duality cascades and $\chi$SB resolution of naked singularities,''
JHEP \textbf{08}, 052 (2000)
doi:10.1088/1126-6708/2000/08/052
[arXiv:hep-th/0007191 [hep-th]].

\bibitem{Greene:1989ya}
B.R.~Greene, A.D.~Shapere, C.~Vafa and S.T.~Yau,
``Stringy cosmic strings and noncompact Calabi-Yau manifolds,''
Nucl. Phys. B \textbf{337}, 1-36 (1990)
doi:10.1016/0550-3213(90) 90248-C

\bibitem{Bergshoeff:1998ry}
E.~Bergshoeff and K.~Behrndt,
``D-instantons and asymptotic geometries,''
Class. Quant. Grav. \textbf{15}, 1801-1813 (1998)
doi:10.1088/0264-9381/15/7/002
[arXiv:hep-th/9803090 [hep-th]].

\bibitem{Ma:2025mvo}
L.~Ma and H.~L\"u,
``Consistent warped \ensuremath{\mathbb{R}}\,\texttimes{}\,T$^{1,1}$ reduction of heterotic supergravity,''
JHEP \textbf{03}, 203 (2025)
doi:10.1007/JHEP03(2025)203
[arXiv:2501.04771 [hep-th]].

\bibitem{Lu:2025nyo}
K.P.~Lu, H.~L\"u and L.~Ma,
``New BPS States from bosonic/heterotic duality,'' to appear in JHEP.
[arXiv:2505.21623 [hep-th]].

\bibitem{Lu:2011zx}
H.~L\"u, C.N.~Pope and Z.L.~Wang,
``Pseudo-supersymmetry, consistent sphere reduction and Killing spinors for the bosonic string,''
Phys. Lett. B \textbf{702}, 442-447 (2011)
doi:10.1016/j.physletb.2011.07.041
[arXiv:1105.6114 [hep-th]].

\bibitem{Liu:2012jra}
H.~Liu, H.~L\"u and Z.L.~Wang,
``Killing Spinors for the bosonic string and the Kaluza-Klein theory with scalar potentials,''
Eur. Phys. J. C \textbf{72}, 1853 (2012)
doi:10.1140/epjc/s10052-011-1853-5
[arXiv:1106.4566 [hep-th]].

\bibitem{Lu:2011ku}
H.~L\"u, C.N.~Pope and Z.L.~Wang,
``Pseudo-supergravity extension of the bosonic string,''
Nucl. Phys. B \textbf{854}, 293-305 (2012)
doi:10.1016/j.nuclphysb.2011.09.002
[arXiv:1106.5794 [hep-th]].


\bibitem{Gibbons:2003gp}
G.W.~Gibbons and C.N.~Pope,
``Consistent $S^2$ Pauli reduction of six-dimensional chiral gauged Einstein-Maxwell supergravity,''
Nucl. Phys. B \textbf{697}, 225-242 (2004)
doi:10.1016/j.nuclphysb.2004.07.016
[arXiv:hep-th/0307052 [hep-th]].

\bibitem{Salam:1984cj}
A.~Salam and E.~Sezgin,
``Chiral compactification on Minkowski$\times S^2$ of $N=2$ Einstein-Maxwell supergravity in six dimensions,''
Phys. Lett. B \textbf{147}, 47 (1984)
doi:10.1016/0370-2693(84)90589-6

\bibitem{Lima:1999dn}
E.~Lima, H.~L\"u, B.A.~Ovrut and C.N.~Pope,
``Instanton moduli and brane creation,''
Nucl. Phys. B \textbf{569} (2000), 247-261
doi:10.1016/S0550-3213(99)00478-2
[arXiv:hep-th/9903001 [hep-th]].

\bibitem{Lima:1999xj}
E.~Lima, H.~L\"u, B.A.~Ovrut and C.N.~Pope,
``Supercharges, Killing spinors and intersecting gauge five-branes,''
Nucl. Phys. B \textbf{572}, 112-130 (2000)
doi:10.1016/S0550-3213(99)00829-9
[arXiv:hep-th/9909184 [hep-th]].

\bibitem{Chamseddine:1999uy}
A.~H.~Chamseddine and W.~A.~Sabra,
``$D = 7$ $SU(2)$ gauged supergravity from $D = 10$ supergravity,''
Phys. Lett. B \textbf{476}, 415-419 (2000)
doi:10.1016/S0370-2693(00)00129-5
[arXiv:hep-th/9911180 [hep-th]].

\bibitem{Cvetic:2000dm}
M.~Cveti\v c, H.~L\"u and C.N.~Pope,
``Consistent Kaluza-Klein sphere reductions,''
Phys. Rev. D \textbf{62}, 064028 (2000)
doi:10.1103/PhysRevD.62.064028
[arXiv:hep-th/0003286 [hep-th]].

\bibitem{Gibbons:1978tef}
G.W.~Gibbons and S.W.~Hawking,
``Gravitational multi-instantons,''
Phys. Lett. B \textbf{78}, 430 (1978)
doi:10.1016/0370-2693(78)90478-1

\bibitem{Hawking:1976jb}
S.W.~Hawking,
``Gravitational instantons,''
Phys. Lett. A \textbf{60}, 81 (1977)
doi:10.1016/0375-9601(77)90386-3









\end{thebibliography}
\end{document}